\documentclass[article]{aastex631}

\newcommand{\logg}{log(\emph{g})}
\graphicspath{ {./images/} }
\usepackage{subfigure}
\usepackage{float}

\begin{document}

\title{Identifying Uncertainties in Stellar Evolution Models Using the Open Cluster M67}

\author[0009-0001-4446-833X]{Susan Byrom}
\affiliation{Department of Physics,
University of Illinois Urbana-Champaign, USA}

\author[0000-0002-4818-7885]{Jamie Tayar}
\affiliation{Department of Astronomy, 
University of Florida, USA}

\begin{abstract} %%abstract 

Stellar age estimates are often calculated by interpolating a star's properties in a grid of models. However, different model grids will give different ages for the same star. We used the open cluster M67 to compare four different model grids: DSEP, GARSTEC, MIST, and YREC \footnote{\href{https://doi.org/10.5281/zenodo.12775242}{10.5281/zenodo.12775242}}. Across all model grids, age estimates for main sequence stars were consistently higher than the accepted age of M67, while age estimates for red giant stars were lower. We compared model-generated age and mass values to external constraints as an additional test of the reliability of each model grid. For stars near solar age and metallicity, we recommend using the DSEP model grid to estimate the ages of main sequence stars and the GARSTEC model grid for red giant stars.
\end{abstract}

%% The AAS Journals uses Unified Astronomy Thesaurus concepts: https://astrothesaurus.org

\section{Introduction} \label{sec:intro}

Stellar ages used in fields such as galactic archaeology and exoplanet evolution are commonly determined by fitting a star to a model grid. A stellar model grid is a set of evolutionary tracks generated by a modeling code at a range of initial masses and metallicities. These tracks predict physical parameters (e.g. luminosity and temperature) of a star as a function of age, initial mass, and initial composition. Therefore, given a model grid and sufficient observational constraints, the age and mass of a star can inferred. However, the assumptions and calibrations that go into creating a stellar model grid can cause significant differences between different grids' age estimates of a star, sometimes more than 30\% \citep{Tayar2022}.

Open star clusters are commonly used to check the accuracy of stellar model grids. In particular, the cluster M67 is often used to calibrate stellar models \citep{Choi2016} because M67 is a well-studied, nearby old open cluster that is approximately 4.0 Gyr old and is near solar metallicity ($[Fe/H]=0.00 \pm .05$) \citep{Myers2022}. 

The age of M67 has been calculated many times. Most commonly, the age of a cluster can be determined by plotting the stars on a color-magnitude diagram and fitting an isochrone to the main sequence turnoff. For M67, Victoria-Regina isochrones give an age estimate of 3.6–4.6 Gyr, with an average of 4.0 Gyr \citep{VB&S2004}. \citet{Sandquist2021} used MIST, PADOVA, and BASTI isochrones constrained by the eclisping binary WOCS 11028 to get an age of 3.5–4.0 Gyr. \citet{Stello2016} used asteroseismology of red giants to obtain an age estimate of 3.46 $\pm$ 0.13 Gyr.

Model grids can be used to estimate the ages of individual M67 stars, essentially treating them as field stars, which are stars that do not belong to a cluster. With perfect models and data, this would return the same age for each star in M67. Systematic discrepancies in the age estimates can reveal flaws in the model grid or input data. We did this for the model grids DSEP, GARSTEC, MIST, and YREC, with model parameters as presented in \citet{Tayar2022}.

The age of M67 has been previously estimated with some of the model grids used in this work. 
\begin{itemize}
  \item MIST isochrones were calibrated using M67 at 4.0 Gyr and solar metallicity  \citep{Choi2016}.
  \item The DSEP model grid isochrone age estimate for M67 is 4.0 Gyr \citep{Dotter2008}.
  \item In \citet{Magic2010}, GARSTEC's age estimate for M67 was determined to be 4.2 Gyr assuming Z/X = 0.0165 and 4.5 Gyr assuming Z/X = 0.0230.
  \item YREC model grids from \citet{VianiBasu2017}, which are slightly different from those used here, estimate the age of M67 at 3.6-4.8 Gyr (best fit at 4.4 Gyr).
\end{itemize}

\section{APOGEE Data} \label{sec:style}

Data from the Apache Point Observatory Galactive Evolution Experiment (APOGEE) Data Release 17 (DR17), as compiled by the Open Cluster Chemical Analysis and Mapping Survey (OCCAM) \citep{Myers2022} lists 663 members of M67, 608 of which have recorded surface gravity (\logg), effective temperature, and metallicity values. We discarded stars with a membership probability of 0 in any of the four categories from OCCAM: proper motion (\textbf{200 stars}), [Fe/H] (\textbf{166 stars}), radial velocity (\textbf{188 stars}), and \citet{Cantat2018} (\textbf{232 stars}). \textbf{Values from \citet{Cantat2018} exist for all stars considered in this step.}
For the remaining 311 stars, we used \texttt{Kiauhoku} \citep{Claytor2020} to fit each star to the model grids based on \logg, [M/H], and T\textsubscript{eff}. We discarded stars that did not fall within the parameter space covered by the model grids. APOGEE DR17 \logg values are less reliable for low-mass dwarfs, so we discarded stars with \logg$ $ $ > 4.5$. This left 141, 140, 143, and 140 star age and mass estimates for DSEP, GARSTEC, MIST, and YREC respectively.

\section{Results} \label{sec:results}

\begin{figure*}[tb]
\begin{minipage}{0.9\textwidth}
\begin{center}

\begin{subfigure} %[b]
    \centering
    \includegraphics[width={0.4\textwidth}, trim={0 0 1cm 1cm},clip]{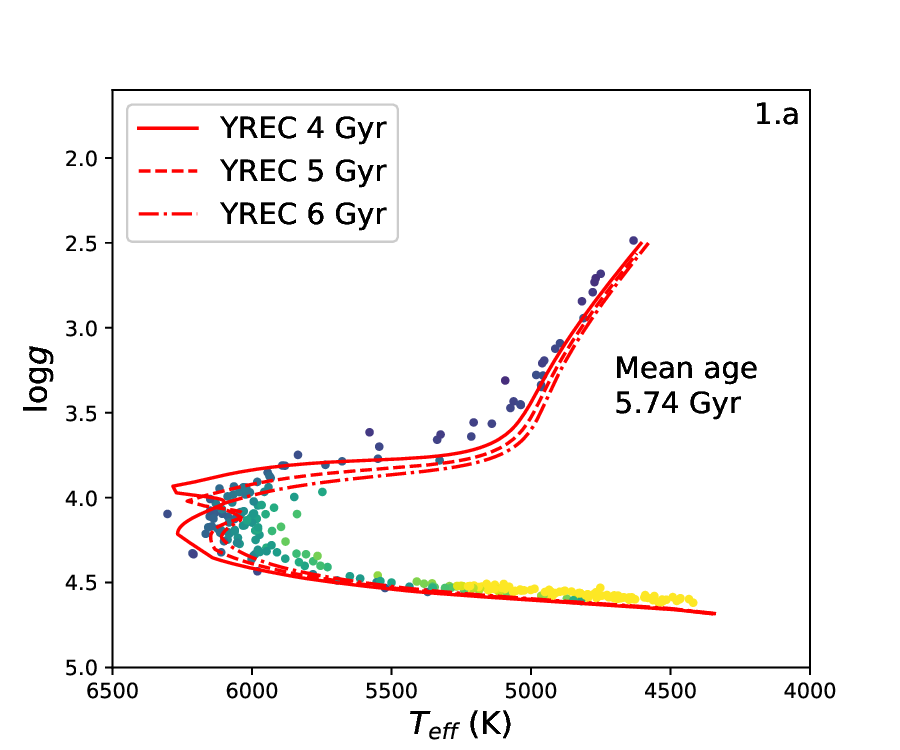}
    \label{fig:yrec}
\end{subfigure}
\begin{subfigure}%[b]
    \centering
    \includegraphics[width={0.47\textwidth}, trim={.2cm 0 0 1cm},clip]{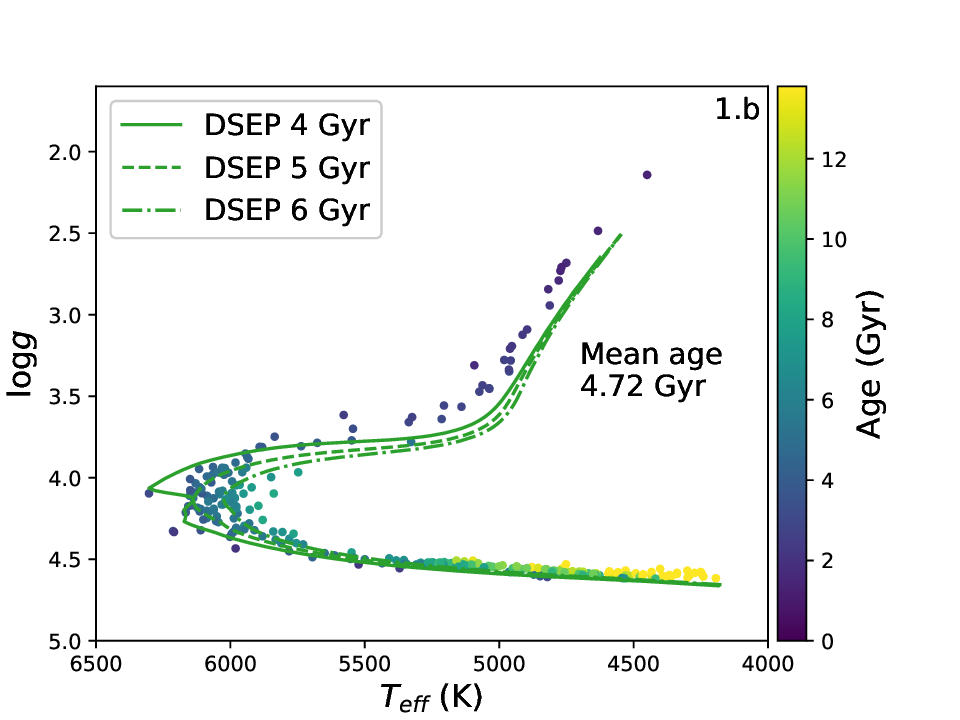}
    \label{fig:dsep}
\end{subfigure}
\begin{subfigure}%[t]
    \centering
    \includegraphics[width={0.4\textwidth}, trim={0 0 1cm 1.4cm}, clip]{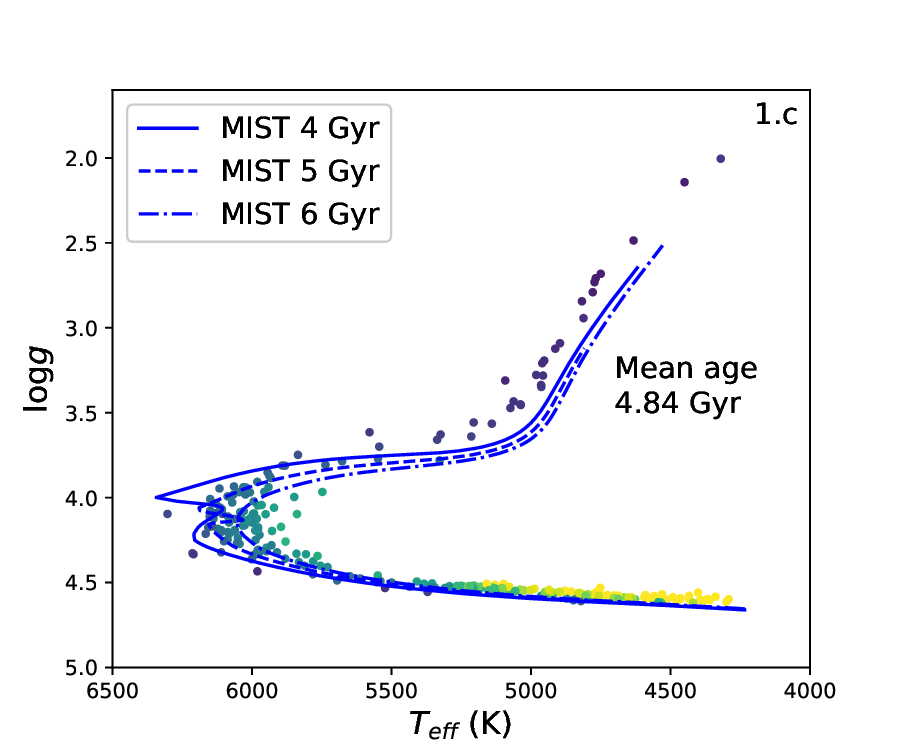}
    \label{fig:mist}
\end{subfigure}
\begin{subfigure}%[t]
    \centering
    \includegraphics[width={0.47\textwidth}, trim={.2cm 0 0 1.4cm},clip]{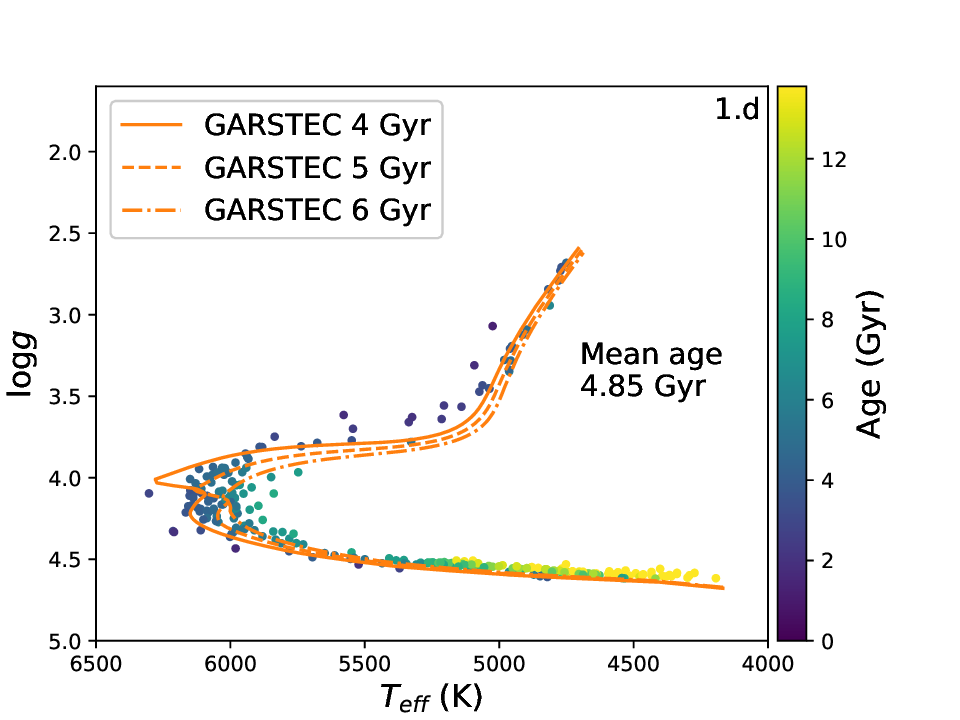}
    \label{fig:gars}
\end{subfigure}
\begin{subfigure}%[t]
    \centering
    \includegraphics[width={0.4\textwidth}, trim={0cm 0cm 1.4cm 1.4cm},clip]{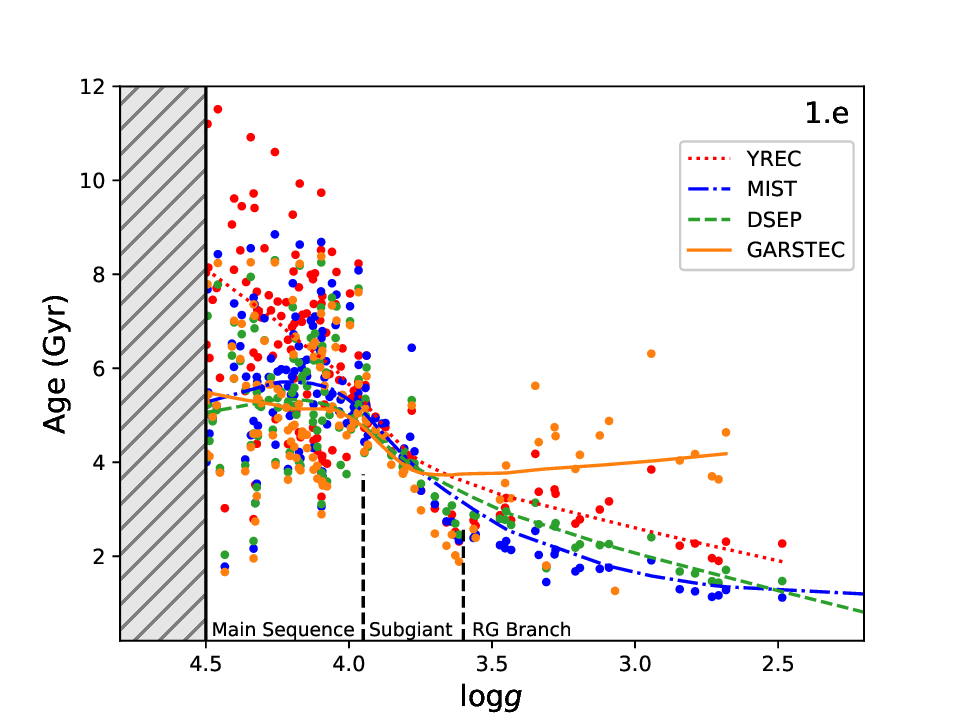}
    \label{fig:agelogg}
\end{subfigure}
\begin{subfigure}%[t]
    \centering
    \includegraphics[width={0.47\textwidth}, trim={.2cm 0cm -1.3cm 1.4cm},clip]{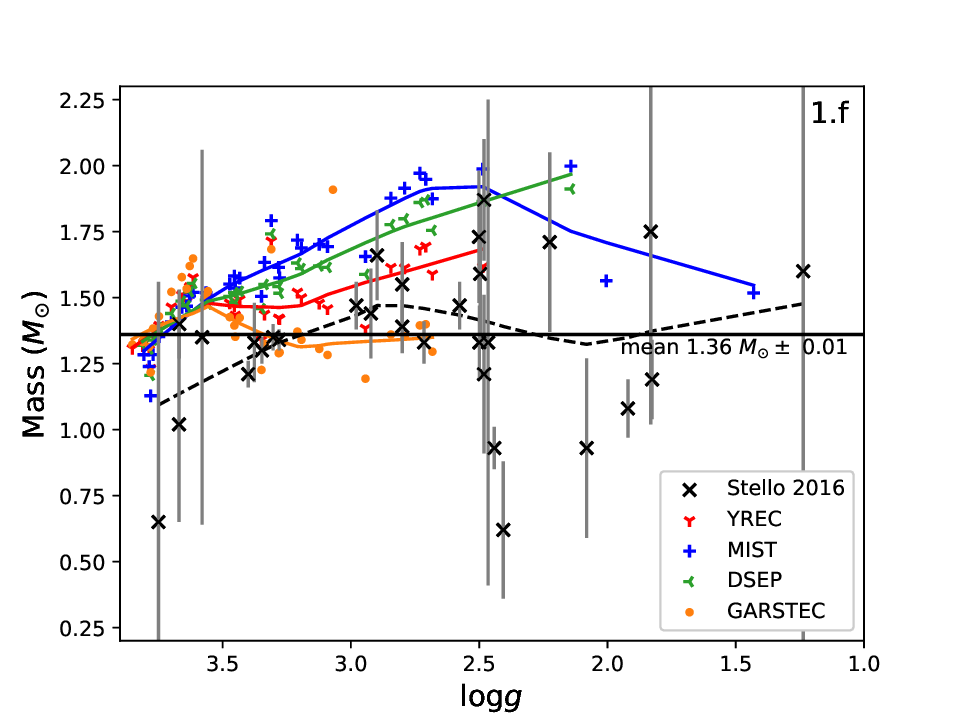}
    \label{fig:rgmass}
\end{subfigure}

\caption{$a$, $b$, $c$, and $d$ are Kiel diagrams of the open cluster M67. Isochrones generated at [M/H] = 0.00 are shown for 4, 5, and 6 Gyr. Stars from M67 were treated as field stars to estimate their ages, which are shown by the colorbar. 1.$e$ shows age as a function of surface gravity for all four models. Vertical dashed lines indicate the approximate boundary between main sequence, subgiant, and red giant stars. The age estimates for each model are fitted with LOWESS non-parametric smooth curves. Notably, the GARSTEC red giant age estimates are near the predicted age of M67, and all models have similar behavior on the subgiant branch. In subfigure $f$, model-generated mass estimates for M67 red giants are compared to \citet{Stello2016} asteroseismic mass estimates for red giants.}
\label{Fig1}
\end{center}
\end{minipage}
\end{figure*}

The average model-generated age of M67 is 4.72, 4.85, 4.84, and 5.74 Gyr for DSEP, GARSTEC, MIST, and YREC respectively. In Figure \ref{Fig1}e, the x-axis is divided into main sequence (MS), subgiant, and red giant evolutionary phases. In Figure \ref{Fig1}a-d, the isochrones are close together on the MS, so a small difference in \logg$ $ between two stars results in a large difference in their estimated ages. We compared ages inferred from the models as a function of \logg$ $ for the models (Figure \ref{Fig1}e). There is a wide spread of MS age estimates, with a model-wide average of 5.03 Gyr. The model-wide average for red giant stars is 2.64 Gyr, and red giant ages from YREC, DSEP, and MIST are lower than the accepted age of M67, while GARSTEC age estimates are more accurate. 

In Figure \ref{Fig1}f, the mass estimates from the models are compared to asteroseismic mass estimates from \citet{Stello2016}. While there were no shared stars between our data and \citet{Stello2016}, we assume that all red giants in a cluster that have evolved as single stars will have similar masses. Of the four model grids, the GARSTEC mass estimates for red giants are closest to the \citet{Stello2016} values. Combined with GARSTEC's reasonable red giant age estimates, this indicates that the GARSTEC-generated model grid is a good choice for estimating parameters of red giant stars near solar age and metallicity.

\section{Conclusions\label{sec:Conclusion}}

When using a model grid to estimate the age of a star, we find that the accuracy of each model grid varies based on the star’s metallicity and evolutionary phase. For stars near solar age and metallicity, the GARSTEC model grid is recommended for estimating the age and mass of red giant stars. For main sequence stars, the DSEP model grid was found to be most accurate. More generally, our findings highlight the need for careful verification and calibration of models in the regime in which they are going to be used.

\begin{acknowledgments}
 We acknowledge support from the National Science Foundation under grant No. 2243878 through the University of Florida 2023 REU. We used data from SDSS (\url{https://www.sdss.org/collaboration/citing-sdss/}).

\end{acknowledgments}

\bibliography{citations}{}
\bibliographystyle{aasjournal}

\end{document}